\documentclass[aip,jcp,reprint]{revtex4-1}
\usepackage{graphicx}
\usepackage{bm}
\usepackage{natbib}
\usepackage{amssymb,amsfonts,amsmath}
\usepackage{CJK}

\begin{document}
\begin{CJK*}{UTF8}{gbsn}
\title{Origins of concentration dependence of waiting times for single-molecule fluorescence binding} 
\author{Jin Yang (杨劲)}
\email[]{Electronic mail: jinyang2004@gmail.com}
\affiliation{Chinese Academy of Sciences --- Max Plank Society
Partner Institute for Computational Biology, Shanghai Institutes for
Biological Sciences, Shanghai 200031, China}

\author{John E. Pearson}
\affiliation{Theoretical Biology and Biophysics, T-10 (Base 6) MS K710, Los Alamos National Laboratory, Los
Alamos, NM 87545, USA}

\date{\today}

\begin{abstract}
Binary fluorescence time series obtained from single-molecule imaging experiments can be used to infer protein binding kinetics, in particular, association and dissociation rate constants from waiting time statistics of fluorescence intensity changes. In many cases, rate constants inferred from fluorescence time series exhibit nonintuitive dependence on ligand concentration. Here we examine several possible mechanistic and technical origins that may induce ligand dependence of rate constants. Using aggregated Markov models, we show under the condition of detailed balance that non-fluorescent bindings and missed events due to transient interactions, instead of conformation fluctuations, may underly the dependence of waiting times and thus apparent rate constants on ligand concentrations. In general, waiting times are rational functions of ligand concentration. The shape of concentration dependence is qualitatively affected by the number of binding sites in the single molecule and is quantitatively tuned by model parameters. We also show that ligand dependence can be caused by non-equilibrium conditions which result in violations of detailed balance and require an energy source. As to a different but significant mechanism, we examine the effect of ambient buffers that can substantially reduce the effective concentration of ligands that interact with the single molecules. To demonstrate the effects by these mechanisms, we applied our results to analyze the concentration dependence in a single-molecule experiment EGFR binding to fluorophore-labeled adaptor protein Grb2 by Morimatsu et al.~\cite{morimatsu2007msr}.
\end{abstract}

\maketitle 
\end{CJK*}

\section{Introduction}
Single-molecule fluorescence techniques measure real-time kinetics of chemical reactions, such as RNA folding~\cite{zhuang2000single}, enzymatic
reactions~\cite{lu1998sme,english2005efs}, protein-protein, protein-oligonucleotides and protein-DNA
bindings~\cite{morimatsu2007msr,teramura2006single,sako2006ism,takagi2011static,wang2009quantitative,elenko2009single}. An important application of single-molecule fluorescence techniques is to probe the binding
kinetics between interaction partners and to infer binding rate constants from the
fluorescence time series. In contrast to conventional ensemble-averaged measurements, single-molecule fluorescence techniques directly observes the binding stochasticity at the molecular level, allowing investigations of conformation fluctuations of individual molecules under various conditions.

Single-molecule fluorescence time series often exhibit transitions alternating between two observable states: fluorescent (on) and non-fluorescent (off) states. One can summarize information in such ``binary" time series using one-dimensional waiting time distributions and kinetic rate constants for association ($k_{\rm on}$) and dissociation ($k_{\rm off}$) can therefore be derived from the mean waiting times. Common experimental procedures usually involve measuring single-molecule binding fluorescence time series under varied ligand concentrations.

These waiting time distributions for protein binding in many cases are fit by sums of multiple exponentials or empirically by stretched
exponentials~\cite{english2005efs,morimatsu2007msr,flomenbom2005sed}, suggesting
that conformation of the single molecule fluctuates during the course of interaction.
Therefore, transitions between the two macroscopic states (``on" and ``off")  proceed through diverse conformation channels connecting microscopic states which cannot be directly distinguished by fluorescence time series. Such temporal variants in protein conformation are referred to as {\it dynamic disorder} if conformations fluctuate on a time scale comparable to that of protein binding~\cite{lu1998sme,zwanzig1990rpd}.

To analyze a binary fluorescence time series, the kinetics of binding between a single molecule and its ligand is usually described by the simple phenomenological two-state model:
\begin{equation}\label{eq:2s}
{\rm off} \displaystyle\underset{k_{\rm off}}{\overset{k_{\rm on}[L]}{\rightleftharpoons}} {\rm
on} \ ,
\end{equation}
where state ``off" indicates that the molecule is free and state ``on" indicates that the molecule is bound to a ligand, and the transition rate from ``off" to ``on" is proportional to the ligand concentration $[L]$ due to the law of mass action.  One can obtain the apparent rate constants $k_{\rm on}$ and $k_{\rm off}$ from the mean waiting times, $\tau_{\rm on}$ and $\tau_{\rm off}$, respectively as:
\begin{equation}\label{eq:rates}
k_{\rm on}=\frac{1}{\tau_{\rm off}[L]}, \ \ k_{\rm off}=\frac{1}{\tau_{\rm on}} \ ,
\end{equation}
where $\tau_{\rm off}$ is the mean waiting time for association (or, the mean dwell time at the macroscopic ``off" state) and $\tau_{\rm on}$ is the mean waiting time for dissociation (or, the mean dwell time at the macroscopic ``on" state). The apparent dissociation constant is then given as $K_d=k_{\rm off}/k_{\rm on}$. In this two-state model, when forward and backward transitions are Markovian, in other words, if the two-state model is biochemically elementary, the rate constants $k_{\rm on}$ and $k_{\rm off}$ are independent of $[L]$. However, measured single-molecule kinetics may potentially be affected by various mechanistic and technical factors including protein concentrations, protein conformations and cellular environments, etc. These factors may cause significant ligand concentration dependence of kinetic rate constants inferred by the mean waiting times as calculated by Eq.[\ref{eq:rates}].

In particular, transitions between observed ``on" and ``off" states can be complex due to conformation fluctuations, such that the two-state minimalistic model of Eq.[\ref{eq:2s}] is non-Markovian and thus is not adequate to describe the true transition mechanism. For such cases, mechanistic models are required to analyze the data.  Here, we examine several possible mechanisms that might cause ligand dependence of the kinetic rate constants $k_{\rm on}$ and $k_{\rm off}$.  Conformation fluctuations in single molecules are described by aggregated Markov models as usually being treated in analysis of single-ion channel recordings~\cite{FREDKINDR:IDEAGG,colquhoun1981stochastic,QinF:Estskp,bruno2005uio}. In addition, we show that non-equilibrium models that violate the detailed balance constraint can also generate strong ligand concentration dependence of rate constants.

We further examine the ligand dependence caused by two technical sources that are not directly related to the binding mechanism: (1) missed events, and (2) background buffer. The former is concerned with transient events that are not captured by observations, which causes overestimation and distortion of ligand dependence in waiting times. We show that missed events have an effect on waiting times similar to that by models of single protein binding site with non-fluorescent binding. The latter mechanism is concerned with ligand interactions with ambient buffer molecules, which causes a substantial reduction in the effective concentration of ligands that bind to the single molecules. Such a background buffering mechanism if unaccounted will cause a strong dependence of the apparent association rate constant on the total ligand concentration $[L]$.

We apply our models to analyze an  {\it in vitro} experiment of single epidermal growth factor receptor (EGFR) binding to adaptor protein Grb2 by Morimatsu {\it et
al.}s~\cite{morimatsu2007msr,takagi2011static}. The experiment showed that waiting time distributions had multiple exponential decay, suggesting that EGFR molecule may have conformational changes on time scales comparable to Grb2 binding. Moreover, the apparent
association rate constant $k_{\rm on}$ had a counter-intuitive dependence on Grb2
concentrations.

\section{Theory}
\subsection{Aggregated Markov models}
The theory of aggregated Markov models was developed for analyzing ion channel gating mechanisms from single ion channel recordings~\cite{FREDKINDR:IDEAGG,colquhoun1981stochastic}. Similar Markov models have been developed more recently in analyzing times series obtained from single-molecule fluorescence experiments~\cite{yang_jpcb2001}. An aggregated Markov model is a special case of hidden Markov chain, in which states in a particular category (an aggregate) correspond to a signal of an identical fluorescence intensity. In a binary single-molecule model, the system fluoresces in states in the ``on" aggregate and does not fluoresce in states in the ``off" aggregate.

An aggregated Markov model for a single molecule can be fully characterized in terms of the ``generator matrix", $Q$, which has an off-diagonal structure that encodes the reaction scheme. Entry $q_{ij}$ is the transition rate from state $i$ to state $j$. The diagonal entries are defined as $q_{ii}=-\sum_{j}q_{ij}$. In a matrix form, one can write ${Qu}=0$, where ${u}$ is the right null vector  of all ones. The
steady-state occupancies ${w}$ is the normalized left null vector of $Q$, i.e.,
${wQ}=0$ and $\sum_iw_i=1$. For systems with aggregates ``on'' and ``off'', $Q$ can be organized and partitioned as
\begin{equation}\label{eq:Q}
{Q}=\left [ \begin{array}{cc} Q_{oo} & Q_{oc} \\ Q_{co} & Q_{cc} \end{array} \right ] \ ,
\end{equation}
where diagonal blocks contain intra-aggregate transition rates and off-diagonal blocks contain inter-aggregate transition rates.  $o$ and $c$ in Eq.[\ref{eq:Q}] denote ``on" and ``off" aggregates, respectively. We can correspondingly partition the null vectors: ${w}=[{w}_o \
{w}_c]$ and ${u}=[{u}_o \ {u}_c]^T$. The mean waiting times are given as
\begin{equation}\label{eq:tauo}
\tau_{\rm on}=\frac{P_{\rm on}}{J_{oc}}, \ \ \tau_{\rm off}=\frac{P_{\rm off}}{J_{co}} \ ,
\end{equation}
each of which is the ratio of steady-state aggregate occupancy, $P_{\rm on}\equiv w_ou_o$ or $P_{\rm off}\equiv w_cu_c$, to the total inter-aggregate probability flux (e.g., $J_{oc}\equiv {w}_oQ_{oc}{u}_c$). The inter-aggregate probability fluxes are balanced at the steady state, i.e., $J_{oc}=J_{co}$.

\subsection{Detailed balance}\label{sec:db}
The principle of microscopic reversibility (or the law of detailed balance)~\cite{yang2006idb} states that at thermodynamic equilibrium for any reversible transition between two neighboring states $i$ and $j$ the probability flux from state $i$ to state $j$ is balanced by that from $j$ to $i$, i.e., $w_iq_{ij}=w_jq_{ji}$. In general, the occupancy $w_i$ has a complex relationship with the ligand concentration $[L]$. However, under the detailed balance, one can derive a relative occupancy $\widetilde{{w}}$, in which each entry has a monomial dependence on $[L]$. This treatment eases analysis of  ligand dependence of $k_{\rm on}$ and $k_{\rm off}$. By the law of mass action, the rate of a ligand binding is proportional to $[L]$, whereas the reversible dissociation rate is a constant. With the detailed balance condition, for two neighboring states $i$ and $j$, we have
\begin{equation}\label{eq:db}
\frac{w_i}{w_j}= \frac{q_{ji}}{q_{ij}}  = K_{ij}[L]^b \ ,
\end{equation}
where $b=-1$ if the state transition from $i$ to $j$ is induced by a ligand binding, $b=1$ if the state transition from $j$ to $i$ is induced by a ligand binding, and $b=0$ if both forward and backward transitions do not involve ligand binding. The coefficient $K_{ij}$ is a constant.  Designate a reference state $r$ at which the molecule does not bind to a ligand. We define the relative occupancy  as the ratio $\widetilde{w}_i\equiv w_i/w_r$, and note that state $i$ and $r$ are connected by a path involving one or more transitions.  Applying Eq.[\ref{eq:db}] successively along a path in the reaction scheme from state $i$
to state $r$, we can show $\widetilde{w}_i= k_i[L]^{n_i}$, where the non-negative integer $n_i$ is the number of ligands bound at state $i$ and $k_i$ is the product of equilibrium constants for all the reversible reactions along the path from state $i$ to state $r$. The numerical value of $k_i$ depends on the choice of the reference state $r$ but does not depend on the choice of a path that connects state $i$ and $r$ due to detailed balance. The mean waiting times are now given as
\begin{equation}\label{eq:tau}
\tau_{\rm on}=\frac{{\widetilde{w}}_o{u}_o}{{\widetilde{w}}_oQ_{oc}{u}_c} \ , \ \ \tau_{\rm off}=\frac{{\widetilde{w}}_c{u}_c}{{\widetilde{w}}_oQ_{oc}{u}_c} \ .
\end{equation}
From the above derivation, it is evident that in general both $\tau_{\rm
on}$ and $\tau_{\rm off}$ are ligand dependent as rational functions of $[L]$.

\begin{figure}
\centering
\includegraphics[scale=0.5]{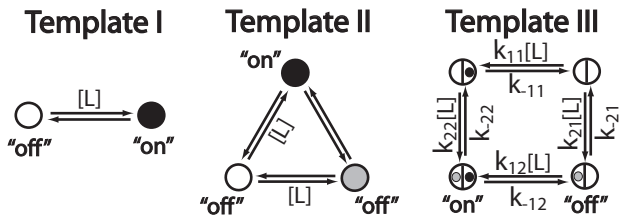}
\caption{\label{fig:scheme} Template I consists of an "on" aggregate in which the receptor has a ligand bound, and an "off" aggregate with no ligand bound. Models constructed from Template I may have any number of ``on" and ``off" states with any connectivity. Rates for transitions from ``off" states (empty circles) to ``on" states (filled circles) are proportional to the ligand concentration $[L]$. All reversible
rates for transitions from ``on" states to ``off" states are constant, and all
intra-aggregate transitions are spontaneous and have constant transition rates.
Template II extends Template I with unresolved ligand-bound ``off" aggregate (gray
circles). Template III describes a single molecule with two binding sites. One
site is fluorescent (dark dot) when binding to a ligand, whereas the other is
nonfluorescent (gray dot). Again there can be any number of states in each of the four aggregates of Template III.}
\end{figure}

\subsection{Model templates} Instead of using models with specific mechanisms, we discuss the apparent
rate constants under different model classes depicted in Fig.~\ref{fig:scheme} by ``template'' reaction schemes. A template consists of several different aggregates. Template I has two aggregates, an ``on" aggregate which has ligand bound and an ``off" aggregate with no ligand bound. Models constructed from Templates I and II can have any number of on and off states and any connectivity. For Template III, the connectivity is arbitrary except that no state in the unliganded aggregate can be directly linked to states in the doubly liganded aggregate. We consider models with two experimentally-distinguishable fluorescent (``on") and non-fluorescent (``off") aggregates with three categories of states for the single molecule; (i) a dark state (``off") with no ligand bound, (ii) a ligand-bound dark state (``off"), and (iii) a bright state (``on") which has ligand bound. The advantage of this approach is that our conclusions do not depend on model details such as the number of states and the connectivity between states but only on key biochemical aspects such as the number of binding sites and whether the binding protein fluoresces while associating with the receptor. Although a waiting time distribution that is a sum of multiple exponents depends on the number of states, the ligand dependence of the apparent association rate does not.

\section{Results}

\subsection{Conformation fluctuations do not cause concentration-dependence of rate constants}
We first consider a single molecule that has a single ligand binding site. We assume that every ligand binding event is experimentally observed, which therefore switches the molecule from an ``off" state to an ``on" state. Each ligand departure switches the molecule from an ``on'' state to an ``off" state. Intra-aggregate state transitions do not involve ligand arrival or departure. Clearly, any such kinetic model can be constructed from a two-state base scheme as shown in Template I (Fig.~\ref{fig:scheme}). Conformation fluctuations could be modeled by extending the base scheme with multiple ``on'' and ``off'' states with an arbitrary connections between states. We can write ${\widetilde w}_o=k_o[L]$ and $\widetilde{w}_c=k_c$, where $k_o$ and $k_c$ are constant vectors. The mean waiting times are given as
\begin{equation}\label{eq:ton}
\tau_{\rm on}=\frac{k_ou_o}{k_oQ_{oc}u_c} \ , \ \ \tau_{\rm off}=\frac{k_cu_c}{[L]k_oQ_{oc}u_c} \ .
\end{equation}
Since $Q_{oc}$ only contains ligand dissociation rate constants $\tau_{\rm on}$ is a constant and $\tau_{\rm off}$ is proportional to the inverse of $[L]$. Thus, we have shown that conformation fluctuations do not generate ligand concentration dependence of the apparent rate constants, $k_{\rm on}$ and $k_{\rm off}$ as defined in Eq.[\ref{eq:rates}].
This result holds for a reaction scheme with arbitrary number of states and arbitrary connections between states as extended from Template I, as long as each binding event was directly observed by the experiment. Notice that in such case $k_{\rm on}$ and $k_{\rm off}$ calculated from the mean waiting times are in fact identical to those obtained in ensemble-averaged measurements.

\subsection{Effect of non-fluorescent binding}  In some cases, ligand binding may not be resolved experimentally and become unnoticed, which could cause ligand dependence as we show below. We consider a model that incorporates non-fluorescent ligand binding to the single molecule. A model (Template II, Fig.~\ref{fig:scheme}) of a molecule that has a single binding site allows a ``dark" conformation (a ``c" state) in which  a bound ligand does not fluoresce (such as due to transient interactions). From Eq.[\ref{eq:tau}], models from Template II give an $[L]$-independent $\tau_{\rm on}$ and a $\tau_{\rm off}$ that has a linear dependence on $[L]$. In particular, one special acyclic scheme from Template II
describes a following two-step binding model.
\begin{equation}\label{eq:diff}
{\rm off}\underset{k_{-1}}{\overset{k_1[L]}{\rightleftharpoons}} {\rm
dark}\underset{k_{-2}}{\overset{k_2}{\rightleftharpoons}}{\rm on} \ ,
\end{equation}
where the first step from the ``off" state to the non-fluorescent ``dark'' state represents a ligand-receptor contact due to ligand diffusion, and the second step to the ``on'' state models a reaction-limited transition. An alternative mechanism is the 2D lateral diffusion of a ligand on the cell surface to search for a binding molecule after a 3D diffusion in the bulk solution onto the cell surface, which may also contribute complications in analyzing the mean waiting times. From the model in Eq.[\ref{eq:diff}], the apparent association rate constant is given as:
\begin{equation}\label{eq:kf1}
k_{\rm on}=\frac{k_1k_2}{k_{-1}+k_2+k_1[L]} \ ,
\end{equation}
where $k_1$ can be considered as the diffusion-limited rate constant. Note that Eq.[\ref{eq:kf1}] contains only two free parameters.

\begin{figure}
\centering
\includegraphics[scale=0.32]{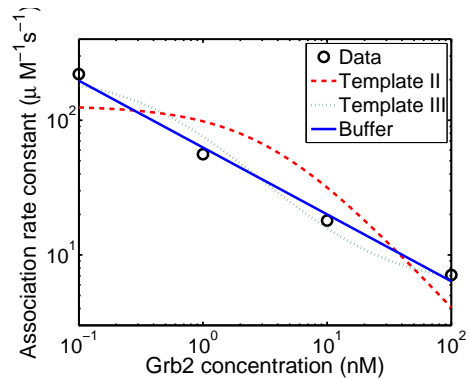}
\caption{\label{fig:nfb} Dependence of the apparent association constant on the
Grb2 concentration. Results are produced by fitting Templates II (dashed curve),
Template III (dotted curve) and the buffering model (solid curve) to the
experimental data (circles) measured by Morimatsu et al.~\cite{morimatsu2007msr}.}
\end{figure}

Morimatsu {\it et al.}~\cite{morimatsu2007msr} hypothesized that frequent and short-term Grb2-EGFR interactions that escaped the instrumental resolution may induce conformation memory in EGFR molecules and thus account for the concentration dependence of the mean off-time. In their experiments, single EGFR molecules were monitored by a total internal reflection fluorescence microscope (TIR-FM) for binding and dissociation with fluorophore (Cy3)-labeled adaptor protein Grb2 that reversibly binds to specific phospho-tyrosine residues on EGFR. Statistics of the fluorescence time series showed that $k_{\rm on}$ decreased from 220 $\mu$M$^{-1}$s$^{-1}$ to 7.1 $\mu$M$^{-1}$s$^{-1}$ when Grb2 concentration increased from 0.1 to 100 nM, whereas the apparent dissociation rate constant $k_{\rm off}$ was found to be constant about 3.4 $s^{-1}$. To apply this model to the data reported by Morimatsu {\it et al}~\cite{morimatsu2007msr}, we assume that the second-order association rate constant is diffusion-limited, $k_1=4\pi Ds=1.51\times 10^3 \mu{M}^{-1}s^{-1}$, where the diffusion coefficient $D=100 \mu$m$^2$s$^{-1}$ and the spherical contact radius $s=2$ nm as in Ref.~\cite{morimatsu2007msr}. The best fit to the data gives $k_{-1}=4.54 s^{-1}$ and $k_2=0.42 s^{-1}$. $k_{-2}$ is measured by the mean on time: $k_{-2}=1/\tau_{\rm on}=3.4 s^{-1}$. As shown by the fitting quality in Fig.~\ref{fig:nfb}, even though $k_{\rm on}$ qualitatively decreases with increasing $[L]$, this scheme is yet to fully quantify the observed dependence on $[L]$, suggesting that an alternative mechanism might better account for the intriguing results reported by Morimatsu {\it et al.}~\cite{morimatsu2007msr}.

\subsection{Molecule with multiple binding sites}
A single molecule with multiple binding sites to a ligand may potentially cause a ligand dependence of the apparent rate constants that have different forms from a molecule with a single binding site. One cause of ligand dependence is that transient ligand binding  escapes observation as discussed below in the missed events section  (Section \ref{subsection: missed}). Here, we assume the existence of a site which does not fluoresce for an unknown reason which does result in different ligand dependence of $\tau_{\rm off}$ than the above model with a single non-fluorescent binding site.

With Template III, we consider a molecule that has two binding sites as shown in Fig.~\ref{fig:scheme}. Although each reversible reaction in Template III involves ligand binding, only one site is monitored for binding. Models constructed from Template III have four state classes: (i) both sites unbound (``off"), (ii) ligand bound to the dark site (``off"), (iii) ligand bound to the bright site (``on"), (iv) ligand bound to both sites (``on").  Calculate the relative occupancy based on the rules described in the previous section (Sec.~\ref{sec:db}). Then, by Eq.[\ref{eq:tau}], the mean on-time is given by
\begin{equation}
\tau_{\rm on}=\frac{k_{-22}+k_{22}[L]}{k_{-11}k_{-22}+k_{-12}k_{22}[L]} \ ,
\end{equation}
which ranges from $1/k_{-11}$ to $1/k_{-12}$ as $[L]$ increases from zero to infinity. When the ligand concentration is small the kinetics of the model is biased to the transitions between the upper two states in Template III, whereas the kinetics is shifted to the transitions between the lower two states when the ligand concentration becomes large. If the off rates $k_{-11}$ and $k_{-12}$ are nearly identical for fluorescent ligand to dissociate from the single molecule bound or unbound to the non-fluorescent ligand, $\tau_{\rm on}$ is independent of $[L]$ and an experiment in this case can only resolve the off rate for the fluorescent binding site. The mean off-time is given by:
\begin{equation}
\tau_{\rm
off}=\frac{k_{-21}k_{-11}k_{-22}+k_{-11}k_{-22}k_{21}[L]}{k_{-21}k_{-11}k_{-22
} k_ { 11 } [ L ] +k_ { -12 } k_{-21}k_ { 11}k_{22}[L]^2} \ .
\end{equation}
Using the above equation and the condition of detailed balance, the apparent association rate constant is given as
\begin{equation}\label{eq:t3}
k_{\rm on}=\frac{k_{11} +k_{12}\frac{k_{21}}{k_{-21}}[L]}{1+\frac{k_{21}}{k_{-21}}[L]} \ .
\end{equation}
As $[L]$ increases from zero to infinity, $k_{\rm on}$ is bounded between $k_{11}$
and $k_{12}$, respectively.  If ligand binding to the fluorescent site is independent of binding to the non-fluorescent site (see Fig.~\ref{fig:scheme}) {\it i.e.} $k_{11}=k_{12}$, then $k_{\rm on}$ does not have ligand dependence.

We use Eq.[\ref{eq:t3}] to fit the data from Morimatsu et al.~\cite{morimatsu2007msr}. In fact, more than one sites on EGFR molecule including phospho-tyrosine sites Y1068 and Y1086 were identified as Grb2 binding sites~\cite{batzer1994hierarchy}. As shown in Fig.~\ref{fig:nfb} (dotted curve), models from Template III with the best fit are able to generate a closer agreement to the data. The fitting gives $k_{11}=2.16\times 10^{2} \mu M^{-1}s^{-1}$, $k_{12}= 5.84 \mu M^{-1}s^{-1}$ and $k_{21}/k_{-21}=2.02\times 10^{3} \mu M^{-1}$.  Since the off-rates $k_{-11}=k_{-12}=3.4 s^{-1}$,  these results indicate a near two orders of magnitude reduction in ligand affinity to the second ligand site after the first ligand binding, suggesting a negative cooperativity of the two binding sites. The parameters also indicate a high affinity non-fluorescent binding with a dissociation constant  $k_{-21}/k_{21}\approx 0.5 nM$.

\begin{figure}
\centering
\includegraphics[scale=0.7]{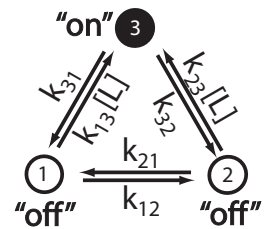}
\caption{\label{fig:db} A three-state cyclic model with two ``off'' states and one ``on'' state. Transitions from ``off'' to ``on'' is induced by ligand binding with transition rates proportional to the ligand concentration $[L]$.}
\end{figure}

\subsection{Effect of detailed balance violation} The above analysis breaks down when the assumption that a system obeys detailed balance becomes invalid, which refers to the situation that a system reaches a non-equilibrium steady state because of reactions driven by an implicit external energy source such as a sustained chemical or electrical potential. The results of Eq.[\ref{eq:tau}] are not applicable when detailed balance does not hold. The steady-state occupancy $w$ must be obtained alternatively (see Appendix).

Here, we first use a minimalistic three-state model (Fig.~\ref{fig:db}) that contains a reaction loop to show that violation of detailed balance causes ligand dependence of rate constants. The model has two ``off'' states without ligand binding and one ``on'' state bound to a ligand. To isolate the effect of violation of detailed balance, the model does not invoke any non-fluorescent ``dark'' state. One can derive the mean waiting times as (see Appendix):
\begin{eqnarray}
\hspace{-0.4in}&&\tau_{\rm on}=\frac{1}{k_{31}+k_{32}}, \\
\hspace{-0.4in}&&\tau_{\rm off}=\frac{(k_{12}+ k_{21})(k_{31}+k_{32})+(k_{23}k_{31}+k_{13}k_{32})[L]}{(k_{31}+k_{32})(k_{12}k_{23} + k_{13}(k_{21}+k_{23}[L]))[L]} \ .
\end{eqnarray}
We note that $\tau_{\rm on}$ is constant regardless the condition of detailed balance. This is a special case for this particular model. In general, violation of detailed balance causes ligand dependence in both mean ``on" and ``off"  times. The association rate constant $k_{\rm on}$ has the same structure as Eq.~\ref{eq:t3} from Template III and would achieve identical best fit to a time series data set as a model constructed from Template III.

Under the condition of detailed balance ($k_{13}k_{32}k_{21}=k_{31}k_{12}k_{21}$), the mean off-time $\tau_{\rm off}$ is reduced to:
\begin{equation}
\tau_{\rm off}=\frac{k_{23}k_{31}+k_{13}k_{32}}{k_{13}k_{23}(k_{31}+k_{32})[L]}\ ,
\end{equation}
which is inversely proportional to $[L]$ and is identical to the result obtained using Eq.[\ref{eq:tau}].

For an arbitrary reaction scheme, we show (see Appendix for detailed derivation using a graph-theoretic method, Fig.~\ref{fig:spanning}) that the apparent rate constants are given as rational functions of $[L]$:
\begin{equation}
k_{\rm on}=\frac{\widetilde{J}}{\widetilde{P}_{\rm off}[L]}, \ \ k_{\rm off}=\frac{\widetilde{J}}{\widetilde{P}_{\rm on}} \ ,
\end{equation}
where $\widetilde{P}_{\rm on}$ and $\widetilde{P}_{\rm off}$ are unnormalized steady-state occupancies of ``on" and ``off" aggregates, respectively. $\widetilde{J}$ is an unnormalized inter-aggregate probability flux. These three terms are all polynomials of ligand concentration $L$. Note that for both $k_{\rm on}$ and $k_{\rm off}$ the denominator and nominator polynomials have corresponding terms of $[L]$ to same powers (See Appendix for mathematical derivation). The exact form of the polynomials is specific to a model topology and the coefficients of the polynomials are in terms of model parameters.  If parameters in a model satisfies detailed balance, in the above equations the ligand-dependent terms factor out from both the nominator and denominator and cancel out, leaving $k_{\rm on}$ and $k_{\rm off}$ ligand independent. The {\it in vitro} experiments by Morimatsu {\it et al.} appear to be done under equilibrium conditions so detailed balance violation is an unlikely  explanation for the observed ligand dependence of $\tau_{\rm off}$.

\subsection{Effect of missed events}\label{subsection: missed}
A missed event is a short-lived binding that escapes the instrumental resolution because it cannot be distinguished from the background noise or because the detector has an intrinsic dead time. Unaccounted missed events distort waiting time distributions and increase mean waiting times. This issue was studied extensively in the field of single ion channel recordings~\cite{colquhoun2009fitting,roux1985general}. Here, we show that missed events may cause the dependence of $k_{\rm on}$ and $k_{\rm off}$ on $[L]$.

Here we analyze the effect of missed events using the two-state model in Eq.[\ref{eq:2s}]. Assume that the measurement has a fixed dead time $\sigma$ and that an event is missed if its waiting time is shorter than $\sigma$. The apparent mean off-time, $\widetilde{\tau}_{\rm off}$, is given as (See Appendix):
\begin{equation}
 \widetilde{\tau}_{\rm off}=\sigma+\sum_{k=0}^\infty q_\sigma
p_\sigma^k[k\tau_\sigma+\tau_{\rm off}(k+1)]=\frac{\tau_{\rm
off}+p_\sigma\tau_{\rm on}}{q_\sigma} \ ,
\end{equation}
where the $\sigma$ accounts for the dead time skipped before the onset of the next detectable on-time interval. The effect of missed events on $\tau_{\rm on}$ is ignorable for small ligand concentrations that only induce less frequent binding. This condition is often satisfied in TIR-FM experiments because binding events should be made rare enough to reduce spatial crowding and the background noise and thus to allow detection in changes in the level of fluorescence signals. This technical requirement limits the concentration at the order of 10nM~\cite{van2011single}. The association rate constant is obtained as
\begin{equation}
k_{\rm on}=\frac{1}{\widetilde{\tau}_{\rm
off}[L]}=\frac{q_\sigma}{1/k_++p_\sigma\tau_{\rm on}[L]} \ .
\end{equation}
This result is mathematically equivalent to that from the single site protein
with non-fluorescent interactions (Eq.[\ref{eq:kf1}]). The model can be derived
from Template II without the transitions between ``on" and ``dark" states:
\begin{equation}
{\rm dark} \ \underset{[L]}{{\rightleftharpoons}} \ {\rm
off} \ {\overset{[L]}{\rightleftharpoons}} {\rm on} \ ,
\end{equation}
where the ligand binding from ``off'' to ``dark'' is used to model the missed events.

We apply the above result to estimate the $k_{\rm on}$ and the relative dead time $\sigma/\tau_{\rm on}$ in the experiment by Morimatsu et al.~\cite{morimatsu2007msr}. From the best fitting shown in Fig.~\ref{fig:nfb} (dashed curve), we identify that $k_{\rm on}=1.15$ nM$^{-1}$s$^{-1}$ and $\sigma/\tau_{\rm on}=2.19$. The dead time $\sigma$ is more than 2 times the mean on-time and the dissociation constant $K_d$ is about 3 nM, suggesting that the affinity between the phosphorylated EGFR and Grb2 is somewhat overestimated, compared to
experimental measurements at 700 nM~\cite{chook1996grb2} and 30nM~\cite{batzer1994hierarchy}. This result reflects the same structural limitation by models from Template II, which did not generate a good fit to the measured data of $k_{\rm on}$.

\subsection{Effect of an external buffer} Here we examine another possible mechanism in which an ambient buffer may sequester ligands (specifically or non-specifically) and consequently reduce the concentration of free ligands available to bind the single molecule of interest. Unaccounted background buffering may cause ligand dependence of $k_{\rm on}$. In the absence of a buffer, the effective ligand concentration $[L]$ that interacts with the single molecule equals the total ligand concentration $[L]_{\rm tot}$. Otherwise, ambient buffers may offset the effective ligand concentration available for binding. We consider a following simple buffering mechanism.
\begin{equation}
 \alpha L+B \overset{K_B}{\rightleftharpoons} L_B \ ,
\end{equation}
where $\alpha\ge 1$ measures the (average) degree of binding cooperativity
between the ligand and the buffer group. In the presence of the buffer, the free
ligand concentration $[L]$ is the effective concentration of ligands that interact with the single-molecule of interest. We note that the specified ambient buffer $B$ can be a mixture of
several kinds of molecules that may interact with the ligand pool. With a
phenomenological equilibrium dissociation constant $K_B$ and the total ligand
concentration $[L]_{\rm tot}$, we have
\begin{equation}\label{eq:buf}
[L]^\alpha[B]=\frac{K_B}{\alpha}([L]_{\rm tot}-[L]) \ .
\end{equation}
Considering that buffer $B$ is in excess ($[B]\gg [L]_{\rm tot}$) so that any changes in  ligand concentration due to binding are insignificant, we have $[L]^\alpha+\beta([L]-[L]_{\rm tot})=0$, where $\beta=K_B/(\alpha [B])$. If the majority of ligands are sequestered, ($[L]\ll [L]_{\rm tot}$), we can approximate the effective ligand concentration $[L]\approx (\beta[L]_{\rm tot})^{1/\alpha}$. Using the two-state model of ligand-receptor binding (Eq.[\ref{eq:2s}]), we obtain a fit (Fig.~\ref{fig:nfb}, solid curve) to the apparent association rate constant data from Morimatsu et al.~\cite{morimatsu2007msr} by $k_{\rm on}=k_+[L]/[L]_{\rm tot}$, and obtained parameter values of $\alpha=1.99$ and $k_+\beta^{1/\alpha}=2.03 \mu M^{-1/\alpha} s^{-1}$. The fitting results suggest that a strong cooperativity ($\alpha\approx 2$) existed for Grb2 binding to the buffer. Although the fitting did not directly resolve $k_+$ and $\beta$, we can make a crude estimation to $\beta\approx 1.8\times 10^{-6} \mu{M}$ (i.e., $K_B/[B]=3.6\times 10^{-6} \mu{M}$) by assuming a diffusion-limited second-order association rate constant $k_+=4\pi Ds=1.51\times 10^{3} \mu{M}^{-1}s^{-1}$ (with $\alpha\approx 2$).

We note that although the above external-buffer model produced the closest agreement (partially due to the mathematical properties of the fitting function) to the data by Morimatsu et
al.~\cite{morimatsu2007msr} in comparison to the previous ones (Fig.~\ref{fig:nfb}), it remains unclear whether the experiment setup introduced a chemical or physical environment that might serve as ambient buffers for Grb2.

\section{Discussion}
Molecular binding is an essential biochemical interaction, which can now be probed at the single-molecule level with fluorescence techniques such as F\"{o}rster (Fluorescence) resonance energy transfer and TIR-FM~\cite{van2011single}. These techniques unveil interaction details that are often unavailable in data obtained from ensemble-averaged experiments. Proper interpretation of the fluorescence time series for single molecule binding by its partner protein (or ligand) requires caution. Especially, phenomenological binding constants $k_{\rm on}$ and $k_{\rm off}$ as well as the dissociation constant $K_d$ extracted from the fluorescence time series may change as the ligand concentration varies, which carries important information about the binding biochemistry and its experimental environment. Model-based analysis of the ligand-dependence of kinetic parameters can help to uncover the underlying mechanisms.

In this paper we explore influences by various mechanistic and technical factors, specifically, single-site and multisite non-fluorescent binding, non-equilibrium steady-states, missed events, and ambient buffers, which could potentially introduce dependence of mean waiting times and thus apparent kinetic rate constants on ligand concentration. A combination of these factors can further obscure the analysis of single-molecule kinetics, requiring assistance of appropriate kinetic models.

We have shown that molecular conformation fluctuation (or dynamic disorder) alone does not cause concentration dependence under the condition of detailed balance in models that reach equilibrium steady states as long as each ligand-induced state transition is experimentally resolved (Template I, Fig.~\ref{fig:scheme}). In this case, kinetic rate constants inferred from mean waiting times reconcile with those measured by ensemble-averaged experiments.

Unobserved ligand binding, due to unknown biochemical reasons, are the essential sources of ligand dependence of the waiting times, which we analyzed using kinetic models that invoke non-fluorescent liganded states. Different models generate different mathematical structures of ligand dependence. In general, a kinetic rate constant, $k_{\rm on}$ or $k_{\rm off}$, is a rational function of ligand concentration.  Models with non-fluorescent liganded states for a molecule that has a single ligand binding site (Template II, Fig.~\ref{fig:scheme}) predict that $k_{\rm on}$ has an inverse linear relationship with ligand concentration $[L]$ (Eq.[\ref{eq:kf1}]), whereas $k_{\rm off}$ remains unmodulated by $[L]$. Models of a molecule with two ligand binding sites with one site non-fluorescent when bound to ligand (Template III, Fig.~\ref{fig:scheme}) predict that both $k_{\rm on}$ and $k_{\rm off}$ have sigmoidal shaped relationship with the ligand concentration.

Unmonitored binding can also be caused by short transitions called missed events whose time durations fall within the length of the dead time of the experimental instrument, which has a similar form of concentration dependence by the single site non-fluorescent binding models (Template II, Fig.~\ref{fig:scheme}). Our results coincide with a similar three-state model proposed by Crouzy and Sigworth~\cite{crouzy1990yet} to account for missed events in single-ion channel recordings, in which transient transitions between a closed state to a short-lived state were used to capture events that was off the scope of the instrumental resolution. It was known in analysis of single-ion channel recordings that unaccounted missed events due to fixed dead time can distort the waiting time distributions and cause overestimation of waiting times. Such limitation may be carried over to cause ligand concentration dependence in single-protein fluorescent binding experiments.

The aforementioned models were studied under the condition of detailed balance. Another source that likely causes ligand dependence is the violation of detailed balance in model parameters, which can be studied using non-equilibrium models. This mechanism is ignored in most studies. The typical assumption of a single-molecule analysis is that the system relaxed to its thermodynamic equilibrium at the steady state. The equilibrium assumption is rather strong and requires the system to meet stringent conditions (the thermodynamics requires the system being isolated without energy and material exchange with its external environment), and it may not be always justified in particular for {\it in vivo} systems that entail many energy-driven reactions~\cite{qian2007phosphorylation} or for {\it in vitro} systems that are sustained by energy sources. Detailed balance violation can be tested by analyzing time series data. For example, two-dimensional joint waiting time distributions that account for two consecutive events, waiting time for binding event followed by that of a dissociation event, can be used to test whether detailed balance holds by checking the time reversibility (also see Ref.~\cite{rothberg2001testing} for other methods). As a consequence of adopting non-equilibrium model, model parameters might not be constrained by detailed balance. A non-equilibrium model achieves a steady state with net fluxes around reaction loops, which gives rise to ligand dependence of rate constants as rational functions of ligand concentration.

We note that our analysis of the effect of detailed balance is closely related to recent works that study the substrate-dependence of enzymatic turnover rate $v$ in fluctuating enzymes with multiple conformation channels~\cite{cao2011michaelis,wu2011}. Under the detailed balance condition, the dependence of production formation velocity on substrate concentration $[S]$ maintains the classic Michaelis-Menten form, $v=\widetilde{k}[S]/(\widetilde{K}_M+[S])$, where effective catalytic rate constant $\widetilde{k}$ and apparent Michaelis-Menten constant $\widetilde{K}_m$ can be derived from kinetic parameters of the model for the enzyme system. When the condition of detailed balance does not hold, $v$ becomes in general a rational function of $[S]$. To demonstrate that the results from our work can also be applied to analyze the turnover rate of multi-conformational enzymes, consider a general scheme of enzymatic network, where an enzyme fluctuates among several ($m$) conformations, forming parallel and interconnected catalytic channels. Through each channel, the enzyme engages the substrate and then undergos multiple ($n$) reversible intermediate steps before finally converting the substrate into a product. The turnover rate can be expressed as the summation of turnover rates in all individual channels: $v=\sum_{i=1}^m\eta_{ni}k_i$, where $\eta_{ni}$ is the steady-state residence probability at the last ($n$th) substrate-bound step of the $i$th channel and $k_i$ is the catalytic rate constant of that channel. We can write $\eta_{ni}=\widetilde{\eta}_{ni}/Z$, where the unnormalized residence probability $\widetilde{\eta}_{ni}$ and the partition function $Z=\sum_{j=1}^n\sum_{i=1}^m\widetilde{\eta}_{ji}$. As shown in Section~\ref{sec:db}, under the detailed balance $\widetilde{\eta}_{ni}$ is proportional to the substrate concentration $[S]$ and $Z$ is a linear function of $[S]$. Thus, the conventional Michaelis-Menten form is preserved in $v$. Without detailed balance, the turnover rate $v$ assumes a rational functional form of $[S]$, which can be obtained systematically using the graphic method as shown in the Appendix C.

Finally we studied the effect by an external buffer group that sequesters ligands, which if unaccounted could cause strong ligand dependence in rate constants. The extent of the buffering effect depends on biochemical nature of ligand-buffer interaction and the relative availability of the buffer group. It is natural to consider buffering in cellular environment of living cells where molecules are subject to ubiquitous binding reactions in a crowded molecular surrounding by specific and/or non-specific interactions.

We applied our results to analyze the experiment data of labeled Grb2 binding to EGFR molecules by Morimatsu et al.~\cite{morimatsu2007msr}. We examined the possibility that missed events due to transient binding (and showed that this is equivalent to Template II) were the source of the ligand dependence of the apparent association rate constant and found that the best fit could not accurately reproduce the data. The mathematically simplest and best fit resulted from assuming there were background Grb2 buffers characterized by two parameters accounting for cooperativity and affinity. Non-equilibrium models with detailed balance violation were not applied to analyze the data because the {\it in vitro} experiments by Morimatsu et al.~\cite{morimatsu2007msr} were apparently performed under the equilibrium condition. Elucidation of the most likely mechanism requires further experimental investigation.

\begin{acknowledgments}
We thank Byron Goldstein, Steven N. Evans, Michael J. O'Donnell and Yandong Yin for helpful discussions. The study was supported by National Science Foundation of China grant 30870477 and Sanofi-SIBS Innovation Grant SA-SIBS-DIG-03 (J.Y.), and by NIH grant R01GM065830-07 (J.E.P).
\end{acknowledgments}

\section*{Appendix}
\renewcommand{\theequation}{A\arabic{equation}}
\setcounter{equation}{0}

\subsection{Aggregated Markov model} An aggregated Markov model of a single molecule kinetics can be described by the following master equation:
\begin{equation}
\frac{dP(t)}{dt}=P(t)Q \ ,
\end{equation}
where entry $p_{ij}$ in matrix $P$ is the probability of being in state $j$ at time $t$ when the system was in state $i$ at $t=0$. Matrix $Q$ is called ``generator matrix". For systems with aggregates ``on'' and ``off'', $Q$ can be organized and partitioned as
\begin{equation}
{Q}=\left [ \begin{array}{cc} Q_{oo} & Q_{oc} \\ Q_{co} & Q_{cc} \end{array} \right ] \ ,
\end{equation}
where diagonal blocks contain intra-aggregate transition rates and off-diagonal blocks contain inter-aggregate transition rates.  Letters $o$ and $c$ denote ``on" and ``off" aggregates, respectively.

The on-time distribution is given
by~\cite{FREDKINDR:IDEAGG,colquhoun1981stochastic}:
\begin{equation}\label{eq:odwell}
f_{\rm on}(t)={\bf\pi}_o e^{Q_{oo}t}Q_{oc}{u}_c \ ,
\end{equation}
where vector $\pi_o$ is the steady-state distribution of ``on" aggregate entry probabilities over the ``on" states, and it is given as the steady-state probability flux into individual ``on'' states from ``off" states normalized by the total probability flux into the ``on'' aggregate: $\pi_o={w}_cQ_{co}/({w}_cQ_{co}{u}_o)$. The mean on-waiting time is calculated as:
\begin{equation}
\tau_{\rm on}=\int_0^\infty tf_{\rm on}(t)dt=\frac{w_ou_o}{w_oQ_{oc}u_c} \ .
\end{equation}
The off-time $\tau_{\rm off}$ is similarly obtained.

\subsection{Violation of detailed balance in the three-state model} Here we derive the mean ``on" and ``off" waiting times for the three-state model shown in Fig. 3 in the main text. If we arrange the states in the order as labeled in the figure, the generator matrix for this model is given by:
\begin{equation}
Q=\left[\begin{array}{ccc} -k_{12}-k_{13}[L] & k_{12} & k_{13}[L] \\
k_{21} & -k_{21}-k_{23}[L] & k_{23}[L] \\
k_{31} & k_{32} & -k_{31}-k_{32}
\end{array}\right]
\end{equation}
One can find the steady-state occupancy (the left null space of $Q$) as:
\begin{equation}
w=\frac{1}{Z}\left[\begin{array}{c}k_{21}k_{31}+k_{21}k_{32}+k_{23}k_{31}[L] \\
k_{12}k_{31}+k_{12}k_{32}+k_{13}k_{32}[L]  \\
(k_{13}k_{21}+k_{12}k_{23})[L]+k_{13}k_{23}[L]^2
\end{array}\right] \ ,
\end{equation}
where $Z$ is the partition function that normalizes $w$, i.e., $Z=\sum_{i=1}^3w_i$. The magnitude of the net flux (regardless of the direction) around the reaction loop can be calculated as:
\begin{equation}
J_n=|w_1k_{12}-w_2k_{21}|=\frac{[L]}{Z}\left|k_{12}k_{23}k_{31}-k_{21}k_{13}k_{32}\right| \ .
\end{equation}
If the model satisfies detailed balance, then $J_n=0$ and the model parameters obey the following constraint:
\begin{equation}
\frac{k_{12}k_{23}k_{31}}{k_{13}k_{32}k_{21}}=1 \ .
\end{equation}
Thus, the equilibrium state probability can be reduced to:
\begin{equation}
w=\frac{1}{Z}\left[\begin{array}{c}k_{23}k_{31} \\
k_{13}k_{32}  \\
k_{13}k_{23}[L]
\end{array}\right] \ .
\end{equation}
The mean "on" and "off" times can be calculated from Eq.[4] in the main text:
\begin{equation}
\tau_{\rm on}=\frac{1}{k_{31}+k_{32}}, \ \ \tau_{\rm off}=\frac{k_{23}k_{31}+k_{13}k_{32}}{k_{13}k_{23}(k_{31}+k_{32})[L]} \ .
\end{equation}
According to $k_{\rm on}=1/{\tau_{\rm off}[L]}, k_{\rm off}=1/{\tau_{\rm on}}$, both apparent association and dissociation rate constants $k_{a}$ and $k_{\rm off}$ are ligand independent. This is consistent with the results we shown in the main text (Detailed balance). We note that in the three-state model the mean ``off" time is also independent of the intra-aggregate transition rates $k_{12}$ and $k_{21}$.

When detailed balance does not hold in the model the mean ``on" time remains unchanged while the mean ``off" time will assume a form as follows:
\begin{equation}
\tau_{\rm off}=\frac{(k_{12}+ k_{21})(k_{31}+k_{32})+(k_{23}k_{31}+k_{13}k_{32})[L]}{(k_{31}+k_{32})(k_{12}k_{23} + k_{13}(k_{21}+k_{23}[L]))[L]} \ ,
\end{equation}
and the apparent association rate constant is:
\begin{equation}
k_{\rm on}=\frac{(k_{31}+k_{32})(k_{12}k_{23} + k_{13}(k_{21}+k_{23}[L]))}{(k_{12}+ k_{21})(k_{31}+k_{32})+(k_{23}k_{31}+k_{13}k_{32})[L]} \ ,
\end{equation}
which has a ligand dependence similar in form to that by models from Template III (Eq.[12]) and may potentially achieve fitting to date with the same quality. In this specific model, the mean ``on" time is a constant and does not have a dependence on ligand concentration. In general, as we shown below that in non-equilibrium models both $\tau_{\rm on}$ and $\tau_{\rm off}$ are ligand dependent.

\subsection{Ligand dependence in a general scheme for single-site binding} \label{app:db}
Obtaining an analytical solution to the steady-state probability distribution $w$ for a general reaction scheme is unwieldy by directly finding the left null space of the generator matrix $Q$. As an alternative, one can obtain $w$ by a known graph-theoretical approach used in non-equilibrium statistical mechanics~\cite{zia2007probability}, which solves for the steady-state distribution for a non-equilibrium system, as we show below.  Here, note that we only consider single-site fluorescent binding and assume that connections between any two states consist of both forward and backward transitions. Below, we first introduce how to use the method to systematically obtain the steady-state probabilities, and then derive the general formula for the ligand dependence of rate constants.

\begin{figure}
\centering
\includegraphics[scale=0.55]{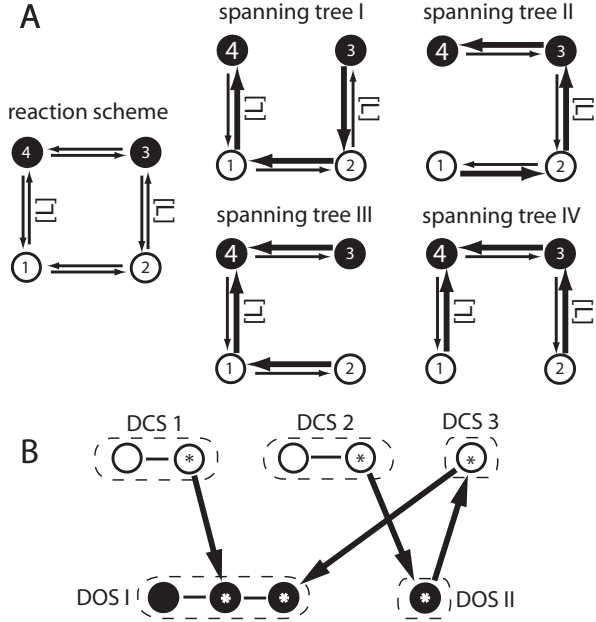}
\caption{\label{fig:spanning} A. A four-state model with 4 distinct spanning trees (I,II,III and IV). Ligand-dependent transitions are labeled with $[L]$. Each of states 3 and 4 has a ligand bound. All links between two states contain a forward transition and a backward transition. In all spanning trees, transitions leading to state 4 are highlighted as thick arrows. B. An illustrative example of a spanning tree that shows 3 disjointed $c$ subtrees (DCS, dashed boxes 1, 2 and 3) connecting to $o$ states in 2 disjointed $o$ subtrees (DOS, dashed boxed, I and II) through gateway states (labeled with $*$'s). The spanning tree can be viewed as a hierarchical acyclic bipartite graph of DCSs and DOSs. The directed edges are shown as directed spanning trees that have root nodes in DOS I. In this spanning tree, the contribution to the unnormalized steady-state probability for an $o$ state is proportional to $[L]^3$, and is proportional to $[L]^2$ for a $c$ state.}
\end{figure}

The method involves enumerating all distinct spanning trees of the topology of a given reaction scheme. A spanning tree of a (undirected) graph is a tree with edges from the original graph that connects all the nodes from the graph. For a topology that has $N$ nodes (states), the maximum number of distinct spanning trees possible is $N^{N-2}$ for the fully connected topology with every pair of nodes directly connected. Fig.~\ref{fig:spanning}A shows all distinct spanning trees for an example four-state model.

For a state $k$, any given undirected spanning tree has a corresponding directed spanning tree $s$ with all unidirectional edges (transitions) leading toward $k$ (see Fig.~\ref{fig:spanning}A for examples). One can view state $k$ as a root of the tree and any directed edge has a direction pointing from an offspring node toward the root. Let $V_{ks}$ be the product of all transition rates associated with the edges in $s$. It is an established result that the steady-state probability for the system to reside in state $k$ is given by~\cite{zia2007probability}:
\begin{equation}\label{eq:ne}
w_k=\frac{1}{Z} \sum_{s}^{N_s}V_{ks}\ ,
\end{equation}
where $N_s$ is the number of distinct spanning trees and $Z\equiv\sum_k^N\sum_s^{N_s}V_{ks}$ is the partition function for normalization purpose. We define the un-normalized steady-state probability vector as:
\begin{equation}
\widetilde{w}=w*Z \ ,
\end{equation}
which we will show has each entry as a polynomial function of ligand concentration $[L]$. We consider an aggregated Markov model of a single molecule binding by a ligand with two aggregates of states, liganded (fluorescent) aggregates and unliganded (non-fluorescent) aggregates.

With the above preparation, we now can show that for the $i$th state of ``on" aggregate in a given directed spanning tree $s$,  $V_{is}$ is a monomial function of $[L]$ with a form $V_{is}=\alpha_{is}[L]^{c_s}$, where integer $c_s$ is the number of disjointed subtrees of $c$ states in $s$.  $\alpha_{is}$ is the product of rate constants of the transitions in $s$. A spanning tree $s$ partitions all $c$ states into $c_s$ ($1\le c_s \le  N_c$) disjointed $c$ subtrees (DCS) (see Fig.~\ref{fig:spanning}B for an example). Each DCS contains only connected $c$ states forming a subnetwork. DCS's have no direct connections to each other but via some $o$ state(s). Each DCS connects to the $o$ aggregate through gateway $c$ states that have direct links with some gateway $o$ states. Similarly, $o$ states in $s$ form several disjointed $o$ subtrees (DOS). For the $i$th state in the $o$ aggregate, according to Eq.[\ref{eq:ne}] the corresponding directed spanning tree provides an additive contribution to the un-normalized steady-state probability $\widetilde{w}_{o_i}$, which consists  one and only one term proportional to $[L]$ by a gateway transition from each DCS such that $V_{is}=\alpha_{is}[L]^{c_s}$. The claim of only one $[L]$-dependent transition from a DCS leading to the $i$th $o$ state is based on the observation that if more than one such transitions exist there will be loop(s) in the spanning tree, which is an obvious contradiction. The result holds for any arbitrary $o$ state in the spanning tree $s$. The result $V_{js}=\beta_{js}[L]^{c_s-1}$ can be derived using similar arguments above for the $j$th $c$ state.

Summing up contributions from all distinct spanning trees for a topology, we then obtain the un-normalized steady-state probability for state $i$ in the $o$ aggregate and state $j$ in the $c$ aggregate:
\begin{equation}
\widetilde{w}_{o_i}=\sum_{s=1}^{N_s}\alpha_{is}[L]^{c_s}, \ \ \widetilde{w}_{c_j}=\sum_{s=1}^{N_s}\beta_{js}[L]^{c_s-1} \ ,
\end{equation}
Therefore, the un-normalized steady-state ``on" and ``off" probabilities are given as:
\begin{equation}
\widetilde{P}_{\rm on}=\sum_{i=1}^{N_o}\sum_{s=1}^{N_s}\alpha_{is}[L]^{c_s},  \ \ \widetilde{P}_{\rm off}=\sum_{j=1}^{N_c}\sum_{s=1}^{N_s}\beta_{js}[L]^{c_s-1} \ .
\end{equation}
The steady-state inter-aggregate flux calculated using the unnormalized ``on" probability is:
\begin{equation}
\widetilde{J}=\sum_{i=1}^{N_o}\sum_{s=1}^{N_s}\gamma_i\alpha_{is}[L]^{c_s} \ ,
\end{equation}
where $\gamma_i$ is the $i$th entry in the vector $Q_{oc}u_c$. Thus, the mean ``on" and ``off" times are given as:
\begin{eqnarray}
\tau_{\rm on}& = & \frac{\widetilde{P}_{\rm on}}{\widetilde{J}}=\frac{\sum_{i=1}^{N_o}\sum_{s=1}^{N_s}\alpha_{is}[L]^{c_s-1}}{\sum_{i=1}^{N_o}\sum_{s=1}^{N_s}\gamma_i\alpha_{is}[L]^{c_s-1}} \ , \\
\tau_{\rm off} & = & \frac{\widetilde{P}_{\rm off}}{\widetilde{J}}=\frac{\sum_{j=1}^{N_c}\sum_{s=1}^{N_s}\beta_{js}[L]^{c_s-1}}{\sum_{i=1}^{N_o}\sum_{s=1}^{N_s}\gamma_i\alpha_{is}[L]^{c_s}}\ .
\end{eqnarray}
The apparent rate constants are:
\begin{eqnarray}
k_{\rm on} & = & \frac{1}{\tau_{\rm off}[L]}= \frac{\sum_{i=1}^{N_o}\sum_{s=1}^{N_s}\gamma_i\alpha_{is}[L]^{c_s-1}}{\sum_{j=1}^{N_c}\sum_{s=1}^{N_s}\beta_{js}[L]^{c_s-1}}\ , \\
k_{\rm off} & = & \frac{1}{\tau_{\rm on}}=\frac{\sum_{i=1}^{N_o}\sum_{s=1}^{N_s}\gamma_i\alpha_{is}[L]^{c_s-1}}{\sum_{i=1}^{N_o}\sum_{s=1}^{N_s}\alpha_{is}[L]^{c_s-1}} \ .
\end{eqnarray}
Therefore, both $k_{\rm on}$ and $k_{\rm off}$ approach to constants at the limits of small and large ligand concentrations. At the intermediate $[L]$, each apparent rate constant is a rational function of $[L]$, whose nominator and denominator have a same structure of a polynomial. With detailed balance, a common $[L]$-dependent term factors from both the nominator and the denominator and the ligand-dependence cancels out. The apparent dissociation constant is given as the following rational function of $[L]$:
\begin{equation}
K_d=\frac{k_{\rm off}}{k_{\rm on}}=\frac{\sum_{j=1}^{N_c}\sum_{s=1}^{N_s}\beta_{js}[L]^{c_s-1}}{\sum_{i=1}^{N_o}\sum_{s=1}^{N_s}\alpha_{is}[L]^{c_s-1}} \ .
\end{equation}

\subsection{Missed event}
Consider a system with a dead time $\sigma$ for detecting binding events. The probability to have a missed binding event is:
\begin{equation}
p_\sigma= \int_0^\sigma \frac{1}{\tau_{\rm on}}e^{-t/\tau_{\rm
on}}dt=1-e^{-\sigma/\tau_{\rm on}} \ .
\end{equation}
Let $q_\sigma=1-p_\sigma$. The mean dead time is calculated as:
\begin{equation}
 \tau_\sigma=\frac{\int_0^\sigma \frac{t}{\tau_{\rm on}}e^{-t/\tau_{\rm
on}}dt}{p_\sigma}=\tau_{\rm on}-\sigma\frac{q_\sigma}{p_\sigma} \ .
\end{equation}
Assuming binding and dissociation events are independent, the probability that one misses $k$ consecutive short events is $p^k_\sigma q_\sigma$. The apparent mean off-time, $\widetilde{\tau}_{\rm off}$, is given as:
\begin{equation}
 \widetilde{\tau}_{\rm off}=\sigma+\sum_{k=0}^\infty q_\sigma
p_\sigma^k[k\tau_\sigma+\tau_{\rm off}(k+1)]=\frac{\tau_{\rm
off}+p_\sigma\tau_{\rm on}}{q_\sigma} \ ,
\end{equation}
where the $\sigma$ accounts for the dead time skipped before the onset of the next detectable on-time interval.

Similarly, the apparent mean on-time is given as
$\widetilde{\tau}_{\rm on}=(\tau_{\rm on}+p_\delta\tau_{\rm off})/q_\delta$,
where $p_\delta=1-e^{-k_+[L]\delta}$ is the probability that an off-time is
shorter than the dead time $\delta$ for detecting a dissociation event. When $[L]$
is very small (i.e., $p_\delta\approx 0$), it is unlikely that a waiting time for a dissociation event falls within the dead time $\delta$ and therefore $\tau_{\rm on}$ is not significantly affected by missed events.

\end{document}